\documentclass[prc,twocolumn,showpacs,preprintnumbers,superscriptaddress,amsmath,amssymb]{revtex4}

\usepackage{amsmath,amssymb,multirow,epsfig,bm}

\newcommand{\beq}{\begin{equation}}
\newcommand{\eeq}{\end{equation}}
\newcommand{\bea}{\begin{eqnarray}}
\newcommand{\eea}{\end{eqnarray}}

\newcommand{\benn}{\begin{displaymath}}
\newcommand{\eenn}{\end{displaymath}}

\begin{document}

\preprint{APS/123-QED}

\title{Tetrahedral correlations in $^{80}$Zr and $^{98}$Zr }

\author{K. Zberecki}, \author{P. Magierski}
\affiliation{Faculty of Physics, Warsaw University of Technology, ul. Koszykowa 75, 00-662 Warsaw, Poland \\}
\author{P.-H. Heenen}
\affiliation{Service de Physique Nucl\'{e}aire Th\'{e}orique, U.L.B - C.P. 229, B 1050 Brussels, Belgium \\}
\author{N. Schunck}
\affiliation{Institute of Theoretical Physics, ul. Hoza 69, 00-631 Warsaw, Poland \\}
\affiliation{University of Surrey, Guildford GU2 7XH, UK \\}

\date{\today}

\begin{abstract}
Axial octupole and tetrahedral correlations in $^{80}$Zr and
$^{98}$Zr have been investigated using the generator coordinate
method, applied to a basis generated by Skyrme HF+BCS calculations.
We focus on the possible presence of states with tetrahedral
symmetry and their stability with respect to octupole vibrations.
We show that pairing significantly reduces the stability of the
tetrahedral configuration and that a shallow mean-field tetrahedral
minimum coexists with an axial octupole minimum. The contributions
to the correlation energies coming from the tetrahedral degree of
freedom and octupole axial deformation are discussed.
\end{abstract}


\maketitle

\section{INTRODUCTION}

The possible tetrahedral instability of finite many-fermion
systems has been predicted in Ref. \cite{hmx}. Recently,
using the similar arguments, it has been conjectured
that the intrinsic ground- or a
low-lying isomeric state of some nuclei may possess a tetrahedral
deformation \cite{ld,dgs}. This kind of symmetry is rather
common in molecules and metallic clusters \cite{kol} where the
mutual geometric arrangement of the ions determines the shape of
the system. In atomic nuclei, the situation is more complicated
and only sufficiently large shell effects can
generate deformed configurations. Although there is convincing
theoretical evidence for the presence of tetrahedral states in
nuclear systems, they have not been confirmed experimentally.

Tetrahedral shapes are invariant with respect
to the transformations of the point group $T_{d}^{D}$. In
nuclei, they are realized at first-order through non-axial
octupole deformations of the nuclear density corresponding to the
intrinsic octupole moment $Q_{32}\propto r^{3}( Y_{3 2} + Y_{3
-2})$.
The key argument in favor of the stability of tetrahedral shapes
is a direct consequence of the theory of point groups. The group
$T_{d}^{D}$ possesses 2 two- and one four-dimensional irrep. The
unusual (in the context of nuclear structure) family of four-fold
degenerate levels lead to a bunching of single-particle states
resulting in rather large shell gaps and increased stability of
specific configurations \cite{dgs,ld,dgs2}

Among the predicted "tetrahedral magic" numbers are the $Z=40$
proton and $N=40, 56-58$ neutron numbers~\cite{dgs}. The Zr
isotopes with 40 and 58 neutrons are thus predicted to be
doubly-magic with respect to this symmetry. However these nuclei
may also exhibit other types of octupole deformations: the
coupling between the neutron $d_{5/2}$ and $h_{11/2}$ orbitals and
the proton $p_{3/2}$ and $g_{9/2}$ orbitals can lead to both axial
and non-axial octupole correlations \cite{bn}. Moreover, octupole
deformations are in competition with the quadrupole mode which may
obscure the signature of the tetrahedral symmetry. Although
experimental data is rather poor for $^{80}$Zr, the rotational
band built on the ground state suggests the presence of a large
quadrupole deformation \cite{fischer}. The evolution of shapes for
the Zr isotopes above the closed shell nucleus $^{90}$Zr is rather
complex. The low energy  spectrum of $^{96}$Zr exhibits a pattern
typical of a spherical nucleus, while $^{100-104}$Zr have very
large quadrupole deformations in their ground state together with
co-existing low-lying oblate and spherical minima \cite{Zr}.
$^{98}$Zr lies at the border between these two regions and
exhibits a transitional character, as demonstrated both
experimentally~\cite{lpk} and theoretically \cite{Zr,hmw,shb93}.
In particular, the large experimental E0 transition between the
first excited $0^{+}$ state and the ground state suggests a strong
mixing between co-existing shapes. This nucleus appears thus to be
particularly rich in terms of the different deformation modes that
are in competition: spherical, oblate, prolate, tetrahedral and
axial-octupole shapes.

The purpose of the present article is to analyze for the first
time the role of tetrahedral configurations in the collective
excitations of these nuclei. It is organized as follows. First, we
investigate the competition between axial and tetrahedral octupole
shapes using the Skyrme Hartree-Fock + BCS (HFBCS) approach to
probe the potential energy landscape in the octupole directions.
We then explore dynamical effects beyond mean-field, parity
restoration and quantum fluctuations, in order to determine
whether states with large tetrahedral correlations are present at low
excitation energy  and thus whether there is a possibility to
identify them experimentally. This analysis, performed using the
generator coordinate method~\cite{gcm}(GCM), aims at paving the
way to a more comprehensive quantum treatment of all quadrupole
and octupole degrees of freedom of the nuclear surface at the
same time. The method has already been applied to the study of
the excitation modes of the superdeformed Hg and Pb isotopes
\cite{shb93a}.

\section{MEAN-FIELD CALCULATIONS}
In order to investigate the variation of the energy of nuclei as a
function of several shape degrees of freedom, the Hartree-Fock
(HF) equations have been solved by discretization on a
3-dimensional mesh in coordinate space~\cite{bfh05}. Contrary to
the usual way of solving mean-field equations in a truncated
oscillator basis, this technique has the advantage that it allows
to describe nuclear configurations with any kind of shape with the
same, high, numerical accuracy. In particular, this method has
been used to describe nuclei from their spherical ground state up
to fission with an accuracy of a few tenth of keV on energy
differences~\cite{bfh05,Hee91}.

Since to study a nucleus as a function of several shape degrees of
freedom is computationally a heavy task, we have imposed a
symmetry condition which simplifies the problem while retaining
the main degrees of freedom relevant for a study of tetrahedral
shapes. The mean-field density is required to be symmetric with
respect to two mutually perpendicular planes. Such a constraint
reduces the complexity of the problem four times but still allows
to study the variation of the nuclear energy as a function of all
octupole degrees of freedom, although odd-$m$ and even-$m$ modes
cannot be treated simultaneously~\cite{shb93a}.
The coupling between multipole moments with even and odd m-values
is beyond the scope of our study, which is focused on the
tetrahedral mode. Note that deformations corresponding to all
multipole moments with even-m values, in particular triaxial quadrupole
deformations are automatically taken into
account by our method.

We have performed a full set of calculations with two
parametrizations of the Skyrme interaction : SIII, which was used in
previous studies of the Zr isotopes with the same
method~\cite{bfh85,shb93}, and SLy4~\cite{SLy}. The results that we
present in details correspond to the SIII force. Both forces predict
very similar behavior of the energy as a function of octupole
degrees of freedom around the spherical configuration. However their
predictions for the energy as a function of the axial quadrupole
moment are qualitatively different. The main difference between both
forces is that, at the mean-field level of approximation, the
$^{80}$Zr ground state obtained with SLy4 is spherical
and the deformed configuration is excited by 3-4 MeV, whereas
for SIII the two configurations are almost degenerate.
Also the spherical configuration of $^{98}$Zr forms a local minimum
with respect to the quadrupole deformation for SLy4 force,
whereas it turns out to be
unstable for SIII.

The pairing interaction has been treated in the BCS
approximation including the Lipkin-Nogami (LN) correction~\cite{ln}. A
zero-range density-dependent pairing interaction has been used:
\begin{equation}
V_{pair}=\frac{1}{2}g_{i}( 1 -P_{\sigma} )
\delta({\bf r}-{\bf r'})\left ( 1 - \frac{\rho({\bf r})}{\rho_{0}}
\right ),
\end{equation}
where $i=n,p$ for neutrons and protons, respectively. As in
previous applications, we set $\rho_{0} = 0.16 \text{fm}^{-3}$.

\begin{table}[h]
\caption{\label{tab:table1} The pairing strengths $g_{i}$ used for
calculations and the averaged pairing gap in the ground state.}
\begin{ruledtabular}
\begin{tabular}{cccccccc}
 SIII  & $g_{p}$ (MeV fm$^{3}$) & $g_{n}$ (MeV fm$^{3}$) & $\bar{\Delta}_{p}$ (MeV) & $\bar{\Delta}_{n}$ (MeV) \\
\hline
$^{80}$Zr & 1100  & 1300  & 1.048 & 1.415  \\
$^{98}$Zr & 1050  & 725   & 0.912 & 0.655
\end{tabular}
\end{ruledtabular}
\vspace*{-0.2cm}
\end{table}
The strengths of the pairing force for both nuclei are listed in
table~I together with pairing gaps $\bar{\Delta}_{i}$ calculated
as an average over the single-particle states within a 5 MeV
window around the Fermi level. These values reproduce the
"experimental" pairing gaps $\Delta$, extracted from the odd-even
mass staggering using a three-point filter from Ref.~\cite{dmn}.
Since there are large uncertainties in the
experimental gaps, especially around $^{80}$Zr, we have also
performed calculations with reduced pairing strengths, in
particular with the values used in Ref. \cite{yam} which give gaps
twice smaller than the "experimental" ones.

\begin{figure}[h]
\vspace*{-0.3cm}
\includegraphics[scale=0.3]{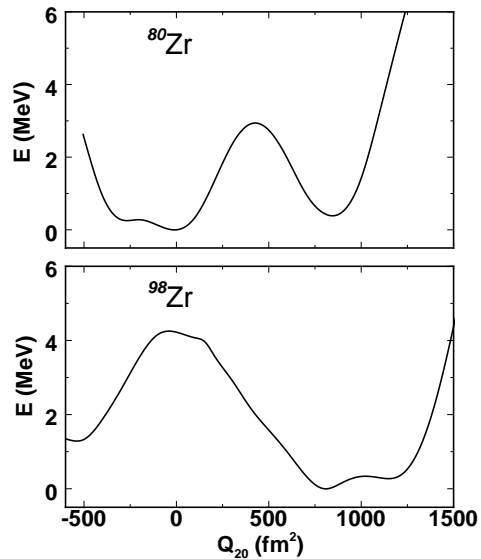}
\vspace*{-1.0cm}
\caption{\label{fig1}Variation of the energy as a function
of the quadrupole $Q_{20}$ moment calculated in the HFBCS
approach.}
\vspace*{-0.4cm}
\end{figure}

The HFBCS+LN calculations presented in Fig.~\ref{fig1} show the
variation of the energy as function of the axial quadrupole
moment. For $^{80}$Zr, the spherical  and deformed
configurations are almost degenerate, while for $^{98}$Zr, the
SIII interaction does not predict a spherical minimum. However
calculations with the SLy4 force give a shallow spherical minimum
with a depth of $200$~keV. These differences between the two
parametrizations are due to the transitional character of $^{98}$Zr
which makes the study of its quadrupole properties
very sensitive to the details of the effective interaction.

\begin{figure}[b]
\vspace*{-1.0cm}
\includegraphics[scale=0.45]{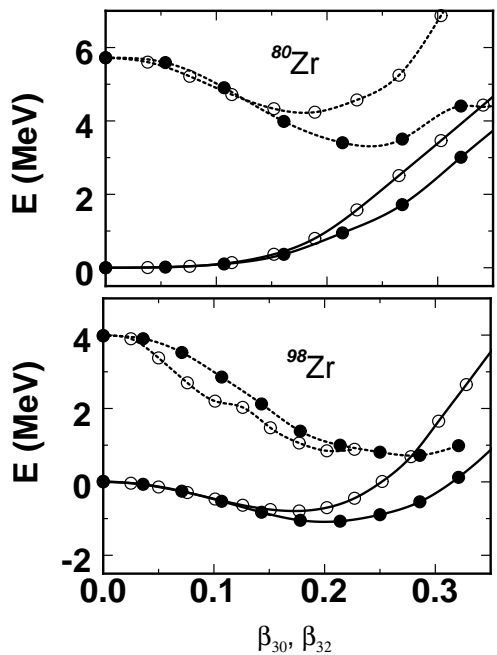}
\vspace*{-0.8cm}
\caption{\label{fig2}Energy as function of the octupole
moments. The quadrupole deformation has been constrained to zero.
The solid and dotted lines denotes the HFBCS results and
pure HF results, respectively. Lines with solid and empty
circles denote the energy as a function of the $Q_{32}$ and
$Q_{30}$ moment, respectively. }
\vspace*{-0.5cm}
\end{figure}

In order to probe the octupole susceptibility of the spherical
configurations, we have performed two sets of calculations along
the axial octupole path ($Q_{32}$ was kept equal to zero) and the
tetrahedral path ($Q_{30}$ was kept equal to zero). In both cases,
the quadrupole moment was constrained to zero. The results are
shown in Fig.~\ref{fig2}. The octupole moments have been expressed
through the dimensionless deformation parameters using the
relation: $\beta_{30}=\langle Q_{30}\rangle /C_{0} $,
$\beta_{32}=\langle Q_{32}\rangle/C_{2} $, where
$C_{0}=\displaystyle{\frac{3}{4 \pi}} A^{2} r_{0}^{3},
C_{2}=C_{0}/\sqrt{2}$ with $r_{0}=1.2 fm$. For each octupole mode,
we have performed calculations with pairing switched off and on.
In the HF approximation, the spherical configuration is always
unstable with respect to the octupole modes. The energy gained by
octupole deformations is around 1~MeV for $Y_{30}$ and is still
larger for $Y_{32}$, around 2~MeV for $^{80}$Zr and 3~MeV for
$^{98}$Zr. This result is consistent with calculations performed
within the macroscopic-microscopic method with a Woods-Saxon
potential ~\cite{dgs}. When pairing correlations are taken into
account, the effects of octupole correlations are strongly
reduced. For $^{98}$Zr the energy gain does not exceed 1~MeV and
for $^{80}$Zr the octupole minima even disappear completely.
Nevertheless the susceptibility towards the $Y_{32}$ mode remains
slightly larger than for the axial octupole mode.
Note that octupole minima in $^{98}$Zr are saddle-points.
Very similar results concerning
the tetrahedral instability of the spherical configuration of
$^{80}$Zr have been found in Skyrme HFBCS~\cite{tak} and
HFB~\cite{yam,polb} calculations. The mechanism by which pairing
correlations makes the tetrahedral minimum weaker is rather
simple. In these nuclei, the single-particle gaps are not large in
the spherical configuration and are increased almost twice by
tetrahedral deformations. As a result, shell effects favor the
creation of a tetrahedral minimum. On the other hand, the pairing
interaction is the largest for the spherical configuration and is
very small in the tetrahedral minimum, resulting in a flat
potential energy curve.

This result shows that a static tetrahedral configuration may
appear only as a result of a delicate balance between pairing and
shell effects, which makes its prediction strongly dependent on
the effective interaction that is used. Moreover should there be a
static tetrahedral minimum, it is likely to be shallow and could
be destroyed by dynamical correlations. One must also emphasize
that our mean-field calculations were performed around the
spherical configuration, by constraining the quadrupole moment to
zero. The octupole minima that are obtained could therefore also
be unstable against quadrupole deformations. Yet, dynamical
factors may also play in favor of octupole minima. In particular,
when octupole correlations are included, the mean-field wave
function breaks parity and parity projection brings a correlation
energy which favors octupole correlation, especially for negative
parity states. We will study these effects in the next section to
determine the stability of a tetrahedral state with respect to
correlations beyond a mean-field approach.

\section{OCTUPOLE DYNAMICS BEYOND A MEAN-FIELD APPROACH}
The generator coordinate method (GCM) allows at the same time to
study the collective dynamics of a nucleus with respect to a
collective variable and to restore the symmetries broken in a pure
mean-field approach (see~\cite{gcm} and references therein). It is
perfectly suited to our goal which is to determine a possible
influence of the tetrahedral mode on low-lying collective states.

The information given by a mean-field potential energy surface
is of a purely static nature and one must go beyond a mean-field
approach to study its stability with respect to vibrations in the
potential well. In cases where the energy surface presents several
co-existing minima separated by low barriers, the very concept of a
mean-field state with a specific deformation may even break down.
Correlations beyond mean-field lead then to states which represent
a mixture of configurations with different deformations. All these
effects are taken into account in the GCM approach.

We will limit here the collective space to octupole modes against
which $^{80}$Zr and $^{98}$Zr are very soft. It is clear that a full
study of these nuclei would require to include also quadrupole
deformations and to mix both octupole modes. Such multi-dimensional
GCM calculations are feasible but represent a heavy task that is
beyond the scope of this exploratory article. The purpose of the
present GCM calculations is limited to find out whether there is a
chance to have a clear signature of the tetrahedral mode in the low
energy spectrum.

\begin{figure}[h]
\vspace*{-0.5cm}
\includegraphics[scale=0.45]{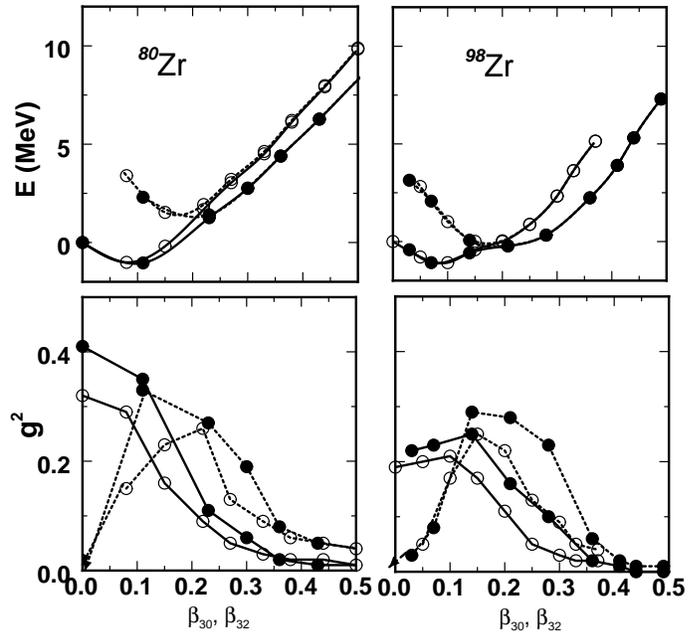}
\caption{\label{fig3} The parity projected mean-field energy as function of
the octupole moments is shown in the upper figures. The solid and
dotted lines denote the positive and negative parity results,
respectively. Lines with open and filled circles denote the
energy as a function of the $Q_{30}$ and $Q_{32}$ moment,
respectively. In the lower figure, the square of collective wave
functions are shown. The line styles correspond to those in
the upper figure.}
\vspace*{-0.2cm}
\end{figure}

We have taken either $Q_{30}$ or $ Q_{32}$ as generator
coordinate.
The wave-functions generated by the constrained mean-field
calculations have first been projected on particle numbers
and parity:
\beq
E(N,Z,\beta_{3\mu})_{\pm} = \frac{\langle
\phi(\beta_{3\mu})|\hat{H}\hat{P}_{(\pm, N,
Z)}|\phi(\beta_{3\mu})\rangle} {\langle
\phi(\beta_{3\mu})|\hat{P}_{(\pm, N,
Z)}|\phi(\beta_{3\mu})\rangle} ,
\eeq
where $|\phi(\beta_{3\mu})\rangle$ are HFBCS wave functions
generated with the constraint $\langle
\phi(\beta_{3\mu})|\hat{Q}_{3\mu}|\phi(\beta_{3\mu})\rangle=C_{\mu}\beta_{3\mu}$.
The operator $\hat{P}_{(\pm,N,Z)}$ is the product of operators
projecting  on $\pi = \pm 1$ parity  and on $N$ neutrons and $Z$
protons. The parity-projected energies are shown in the upper part
of Fig.~\ref{fig3}. As usual after parity
restoration~\cite{shb93a} the energy minima for positive parity
states are shifted towards smaller octupole deformations compared
to the HFBCS minima (see Fig.\ref{fig2}) while the negative parity
states have larger deformations. For both nuclei the energy minima
for positive parity correspond to very similar $\beta_{30}$ and
$\beta_{32}$ values. For the negative parity curve, $\beta_{32}$
is systematically larger than $\beta_{30}$ in the minimum.
Qualitatively  similar results have been obtained with the SLy4
Skyrme parametrization.

The GCM allows to study the stability of the configurations,
corresponding to the minima of these energy curves,
with respect to large amplitude vibrations. A
collective wave function is constructed by mixing the mean-field
states corresponding to different values of the octupole
moment, after their projection on particle number and parity:
\beq
|\Psi\rangle = \int f(\beta_{3\mu}) \hat{P}_{(\pm, N,
Z)}|\phi(\beta_{3\mu})\rangle d \beta_{3\mu}
\eeq
The coefficients $f(\beta_{3\mu})$ are determined by minimization
of the total energy of the collective wave function
$|\Psi\rangle$. The same effective interactions as in the
mean-field calculations are used. In practice, the integral is
replaced by a discrete summation over $\beta_{3\mu}$, with a
number of points large enough to obtain results independent of the
discretization\cite{gcm}. The discretized Hill-Wheeler (HW)
equation was solved separately for each collective coordinate
$Q_{30}$ and $Q_{32}$. The collective wave functions (related to
$f(\beta_{3\mu})$ by an integral transformation) are plotted in
Fig.~3. One can see that these wave functions are spread around
the minima of the projected mean-field energy curves, with a shape
typical of a vibration in a 1-dimensional energy well.

\begin{table}
\caption{\label{tab:table2}Results of the GCM calculations. $E_{exc}$ denotes
          the excitation energy of the first negative-parity collective state
      with respect to the first positive-parity state. $E_{corr}$ is the
      correlation energy as defined in (\ref{CorrEner}) and
      $\tilde{\beta}_{3\mu}$ refers to the dynamical deformation of
      (\ref{DynDef}).}
\begin{ruledtabular}
\begin{tabular}{cccccccc}
 SIII &$E_{exc}$ (MeV)& $E_{corr}$ (MeV) & $\pi$ & $\tilde{\beta}_{30}$ & $\tilde{\beta}_{32}$ \\
\hline
$^{80}$Zr& 0.0   & 1.499 & $+1$ & 0.0  & 0.12 \\
         & 2.22  &  -    & $-1$ & 0.0  & 0.25 \\
         & 0.0   & 1.459 & $+1$ & 0.13 & 0.0  \\
         & 2.52  &  -    & $-1$ & 0.23 & 0.0  \\
\hline
$^{98}$Zr& 0.0   & 1.485 & $+1$ & 0.0  & 0.14 \\
         & 0.784 &  -    & $-1$ & 0.0  & 0.21 \\
         & 0.0   & 1.387 & $+1$ & 0.11 & 0.0  \\
         & 1.09  &  -    & $-1$ & 0.19 & 0.0  \\
\end{tabular}
\end{ruledtabular}
\vspace*{-0.2cm}
\end{table}

The energy gain due to correlations beyond the mean-field is shown in table
II. The correlation energy is defined by:
\beq
E_{corr} = E (N,Z,spher.) - E{^+},
\label{CorrEner}
\eeq
where $E (N,Z,spher.)$ is the energy of the
particle number projected spherical
configuration obtained in the HFBCS approach, and $E{^+}$ is the
lowest positive-parity energy obtained in the GCM. We calculated also
a dynamical deformation associated with the lowest collective
eigenstates for both parities defined by:
\beq
\tilde{\beta}_{3\mu}=\sum_{\beta_{3\mu}} \beta_{3\mu} g_{\pm}^{2}(\beta_{3\mu} ).
\label{DynDef}
\eeq

The correlation energies are very similar for both modes and both
nuclei with however a slightly larger gain for the $Q_{32}$ mode.
This small difference induces also an increase of dynamical
deformations with respect to static ones. More significant are the
differences obtained for the negative parity states whose excitations
are lower by 300~keV. Qualitatively similar predictions are obtained
with the SLy4 interaction.

In order to check the sensitivity of the GCM results on the
magnitude of pairing correlations, we have performed calculations
with pairing strengths that produce gaps larger or smaller
by a factor 2. Qualitatively, the GCM results are not affected: the
changes in the collective wave functions and dynamical
deformations are marginal. The correlation energies vary by
$10-20$\%. A doubling of the pairing gap results in approximately
twice larger excitation energy for the negative parity states. We
have found also that the results for $^{98}$Zr are less sensitive
to the pairing strength than for $^{80}$Zr.

\section{CONCLUSIONS}

We have studied two nuclei, $^{80}$Zr and $^{98}$Zr, which are
doubly magic with respect to the tetrahedral symmetry, in order to
establish whether they may serve as good candidates to obtain experimental
evidence of a stable tetrahedral deformation. Our results, based on the
mean-field approach, indicate that the spherical configuration of
both nuclei is indeed unstable against octupole deformations. The
presence of a tetrahedral minimum is the result of a delicate
balance between shell effects and pairing. Due to the very shallow
energy minima, quantum fluctuations beyond mean-field play a
significant role. In order to quantify this role, we have
performed dynamical calculations using the GCM in two octupole
directions, specified by either tetrahedral or axial octupole
deformations. The correlation energies associated with the
octupole collective modes are large in both cases and lower
significantly the energy of the spherical configuration. These
results seem to be qualitatively independent of the Skyrme parametrization.

We have performed the calculation in such conditions that each mode
is clearly separated from the other, constraining in particular the
quadrupole deformations to be zero. Under these conditions it turns
out that the correlation energies are slightly larger for the
tetrahedral mode than for the axial one. Also the tetrahedral
vibration has smaller energy than the corresponding energy of axial
octupole collective mode. Hence it was shown that thanks to
dynamical effects, the tetrahedral mode is energetically more
favorable as compared to the axial octupole mode. However the
possible mixture between these modes cannot be excluded.

In summary, our results suggest that both the axial and
tetrahedral type of octupole correlations play an important role
in these nuclei, and they possess very similar characteristics.

\begin{acknowledgments}
Discussions with J. Dobaczewski, J. Skalski, W. Satu{\l}a
and P. Olbratowski are gratefully acknowledged. This work
has been supported in part by the Polish Committee for Scientific
Research (KBN) under Contract No.~1~P03B~059~27, the
Foundation for Polish Science (FNP), the \mbox{PAI-P5-07} of the Belgian Office
for Scientific Policy and the Department of Energy under grant DE-FG03-97ER41014.
Numerical calculations were performed at the Interdisciplinary Centre
for Mathematical and Computational Modelling (ICM) at Warsaw University.
\end{acknowledgments}

\end{document}